\documentstyle[preprint,aps]{revtex}
\tightenlines

\begin{document}

\draft

\title{ {\bf Effects of Backflow Correlation in the Three-Dimensional \\
Electron Gas: Quantum Monte Carlo Study}}

\vspace{-0.5cm}

\author{Yongkyung Kwon}
\address{Department of Physics and Center for Advanced Materials and Devices,\\
Kon-Kuk University, Seoul 143-701, Korea}
\vspace{-0.5cm}
\author{D. M. Ceperley}
\address{Department of Physics and National Computational Science Alliance, \\
University of Illinois, Urbana, Illinois 61801}
\vspace{-0.5cm}
\author{Richard M. Martin}
\address{Department of Physics and Materials Research Laboratory, \\
University of Illinois, Urbana, Illinois 61801}

\date{\today}
\vspace{-1.5cm}
\maketitle

\begin{abstract}
The correlation energy of the homogeneous
three-dimensional interacting electron gas
is calculated using the variational and fixed-node
diffusion Monte Carlo methods, with trial
functions that include backflow and three-body correlations.
In the high density regime ($r_s \le 5$) the effects of
backflow dominate over those due to three-body correlations,
but the relative importance of the latter increases
as the density decreases.  Since the backflow correlations
vary the nodes of the trial function, this leads to
improved energies in the fixed-node diffusion Monte Carlo calculations.
The effects are comparable to those found for the
two-dimensional electron gas, leading to much improved
variational energies and fixed-node diffusion energies
equal to the release-node energies of Ceperley and Alder
within statistical and systematic errors.

\end{abstract}


\pagebreak
\narrowtext

\section{INTRODUCTION}
\label{sec:intro}

The homogeneous electron gas in three dimensions is the
simplest model to study the effects of correlation between electrons
in metals\cite{pines}. 
Its correlation energy, defined as the total ground-state energy
minus the Hartree-Fock energy,   
has been used to give the exchange-correlation potential in 
density functional calculations with 
the Local Density Approximation \cite{koh65,par89}. 
The possible phases which this simple system can
display are prototypes for understanding 
interacting electrons in extended matter\cite{bloch,wigner,sdw,pair,mahan}.

The theoretical study of the interacting electron gas
began with Bloch \cite{bloch}
who discovered by using the Hartree-Fock approximation that
the system would favor
a ferromagnetic liquid state over the normal paramagnetic
state at low electron densities. 
Wigner\cite{wigner} first calculated 
the correlation energy of the homogeneous electrons
at high density limit, using the second-order perturbation theory.
He also pointed out that for sufficiently low densities
the electrons would become localized and form an ordered array.
After calculating the correlation energy of this electron solid
with the Wigner-Seitz approximation \cite{pines,mahan},
he proposed an interpolation formula for the correlation energy
in a wide range of densities having his high- and low-density limits.
The development of the field-theoretic approaches in 1950s led to
various approximate methods \cite{mahan} 
to calculate the ground-state properties
of the electron gas. Among them, 
Gell-Mann and Brueckner \cite{gell} summed the ring diagrams
to compute the correlation energy in the high-density limit.
The dielectric function formalism, especially with the self-consistent
treatment of the screening process introduced by Singwi, Tosi, Land, and
Sj\"{o}lander \cite{stls}, gave more accurate ground-state properties
at a wider range of densities.  

On the other hand, various stochastic numerical methods known 
collectively as quantum Monte Carlo (QMC)
have been developed to compute the properties of a quantum many-body
system such as the electron gas.
Ceperley \cite{cep78} first applied variational Monte Carlo (VMC) 
to calculate a much more accurate upper bound to the ground-state energy 
of the electron gas than is given by Hartree-Fock.
More accurate correlation energies were computed 
by Ceperley and Alder \cite{alder} with the
diffusion Monte Carlo (DMC) method
which projects the true ground state of a many-body system from a
trial state. 
Even though the DMC method gives the exact ground-state energy 
for a system of many bosons,
it has a serious difficulty in treating fermion systems, because  
fermion wave functions must be antisymmetric
under particle exchanges \cite{sign}.
In order to 
address this problem, Ceperley and Alder \cite{alder} developed
the released-node method. Its only limitation
is that the statistical fluctuations
can grow rapidly at large  projection time. 
So the statistical noise can dominate the signal before converging to the
ground state. 

In this work we use the fixed-node method \cite{anderson,alder,reynolds82}, 
where
the nodal surface of the exact ground-state wave function is approximated
by that of the trial wave function. 
We adopt the approach of systematically improving the fixed-node DMC results 
by using a trial function with better nodes, analogous
to our work on the two dimensional electron gas \cite{kwon93}.
Unlike the released-node method this method is stable and 
does not have the convergence problem. It gives the best upper bound
to the exact energy consistent with the assumed nodes.

Ceperley and Alder used the Slater-Jastrow trial wave function
in both released-node \cite{alder} and fixed-node calculations \cite{conf}, 
which consists of
the Slater determinant of single-body orbitals and products
of two-body correlation functions.
According to their released-node calculation, the electron
gas could  exhibit three different phases at zero temperature, 
the paramagnetic and ferromagnetic liquids and the Wigner crystal,
depending on its density. 
More recently, Ortiz and Ballone \cite{ortiz} reported  a new
fixed-node DMC calculations with the Slater-Jastrow wave function.
Their correlation energies were found to be,
as expected, 
smaller in magnitude
than Ceperley and Alder's released-node calculations,
especially at high metallic densities.
In this paper, we perform a fixed-node DMC calculation
using a trial function with
backflow and three-body correlations in addition to two-body correlation.
Our calculations in the two-dimensional electron gas \cite{kwon93} showed that
the inclusion of the backflow correlation in
a trial state greatly improved the Slater-Jastrow
fixed-node results. 
We found that, although the Slater-Jastrow wave function accounts for
most of the 
correlation energy of the electron gas, the remaining errors that are in the
Slater-Jastrow function are mostly due to backflow 
and three-body correlations.
This is similar to what has been observed on calculations of the other
strongly correlated system of fermions, liquid $^3$He\cite{Moroni}.
Our fixed-node DMC results in three dimensions 
will be compared with Ceperley
and Alder's released-node results.

All ground-state properties of the electron gas at zero magnetic fields 
are determined only by the dimensionless density parameter $r_{s}=a/a_{0}$,
where $a_{0}$ is the Bohr radius,
$a=(\frac{3}{4\pi \rho})^{1/3}$ is the radius of a sphere which encloses
one electron on the average and $\rho$ is the number density.
With energy units of $Rydbergs$ (Ry) and the length units of $a$,
the Hamiltonian of the electron gas is
 \begin{equation}
    H=- \frac{1}{r_{s}^{2}} \ \sum_{i=1}^{N} \nabla_{i}^{2} + \frac{2}{r_{s}} \
\sum_{i<j} \frac{1}{|{\bf r}_{i}-{\bf r}_{j}|} \ + \
  constant \: ,
 \end{equation}
where the $constant$ is the term due to the uniform background of
opposite charge.  
We consider the density range of $1 \le r_s \le 20$, where
Ceperley and Alder found the system in the normal liquid phase.
We do not consider spin polarized  or superconducting states.

\section{Methodology}

In a VMC calculation, one estimates the properties of a quantum
state, by assuming a trial wave function $ \Psi_{T}(R) $ with the correct
symmetry, where $R=({\bf r}_1,{\bf r}_2, \cdots, {\bf r}_N )$ 
is a $3N$-dimensional vector representing the positions
of $N$ particles. With a set of configurations $\{ R_i \}$ sampled 
with a probability density proportional to $\Psi_T^2(R)$,
the variational energy is just the average of local energies,
$E_{L}(R_i) = H \Psi_{T}(R_i) / \Psi_{T}(R_i)$.
This method can give a good upper bound to the exact energy 
if the trial state is accurate as it is for the
homogeneous electron gas \cite{cep78}.

Even more accurate ground-state properties of 
a many-body system can be obtained
with the DMC method, where the Schr\"{o}dinger equation 
is solved by treating it as a diffusion equation \cite{alder}.
The solution of the Schr\"{o}dinger equation in imaginary time $t$, 
$-\partial |\Phi\rangle / \partial t = (\hat{H}-E_T) |\Phi \rangle$,
can be expressed in terms of the exact energy eigenvalues $E_i$ and
eigenstates $\phi_i$:
\begin{equation}
  \Phi(R,t) = \sum_{i} c_{i} \exp[-t(E_i-E_T)] \phi_i(R) .
\label{eq:tsolution}
\end{equation}
At sufficiently long times only the ground state $\phi_0$
survives in Eq. (2), if $\Phi(R,0)$ is not orthogonal to it.
In order to implement this idea with a stochastic procedure,
we consider the real-space representation of the Schr\"{o}dinger equation: 
 \begin{equation}
    \frac{\partial f(R,t)}{\partial t} = \frac{1}{r_{s}^{2}}\sum_{i=1}^{N} 
   \nabla_{i} \cdot
   (\nabla_{i}f-f\nabla_{i}\ln\Psi_{T}^{2})-(E_{L}(R)-E_{T})f \; ,
 \label{eq:diffeq}
 \end{equation}
where $f(R,t)=\Phi(R,t)\Psi_{T}(R)$.
Note that the Schr\"{o}dinger equation is multiplied by
the trial wave function $\Psi_{T}(R)$.
Eq. (3) can be viewed as a diffusion equation
in a $3N$-dimensional space with the density of diffusing particles
$f(R,t)$. Its second term imposes a drift 
and the final term gives rise to a
branching process by which the
sampled configurations converge to the lowest-energy state. 
The initial ensemble of configurations $\{ R \}$ with probability density
$f(R,0)=\Psi_T^2(R)$ is evolved forward in time by the
above diffusion equation and reaches the equilibrium distribution
$f(R,\infty)=\phi_0(R)\Psi_{T}(R)$
at large enough $t$.  From this distribution of random walks, the exact 
ground-state energy 
$E_0=\langle \phi_0 | \hat{H} | \Psi_T \rangle / \langle \phi_0 | 
\Psi_T \rangle$ can be
estimated as the average of the local energy:
$E_{L}(R) = H \Psi_{T}(R) / \Psi_{T}(R)$.

The diffusion equation formulation described above requires
for implementation
that the population density $f(R,t)$ be non-negative.
For Bose systems, this is not
a problem since their ground-state wave functions can be chosen to be 
non-negative.
However, fermion wave functions are antisymmetric, change sign, and have
nodes.
This leads to the famous {\em sign problem} \cite{sign} 
in the QMC calculations of Fermi systems.
The apparent limitation of the diffusion analogy in this case
can be dealt with by treating positive and negative regions
separately. One easy way to accomplish this is not to allow diffusion
between these two regions, which corresponds
to the $fixed$-$node$ approximation \cite{anderson,alder}.
If $\Psi_T$ were to have the exact nodes of the ground state, one could
treat the fermion system immediately and exactly, since $f$ would never
change sign. Unfortunately, the exact location
of the nodes in many-fermion systems is not known\cite{node}.
The fixed-node approximation is based on the
requirement that $\phi_0(R)\Psi_T(R)$ be non-negative.
The fixed-node DMC energy is an upper bound to the exact energy;
the best upper bound with the given nodes,
and usually lies well below the variational energy \cite{node}.

Another way to deal with the sign problem is to use the
released-node method \cite{alder} which puts no constraints on the
nodal structure of the true ground-state wave function.
In this method, there is a population
of positive random walks which give positive contributions
to any average, and a population of negative walks with negative contributions. 
Whenever a random walk diffuses across the nodes of the trial function,
the sign of its contribution changes.
Even though it can be shown
that the difference population converges to the antisymmetric
fermion ground state, it does not proceed without problems. 
Since both the positive and negative populations grow geometrically
with a large number of projections,
the statistical fluctuations in the average increase
exponentially \cite{alder}. So for this method to be successful
the diffusion process needs to converge
to the ground state before the fluctuations become large.
As the system size gets larger, the fluctuations grow. 
Hence, the fixed-node method is more useful for systems with many fermions.

\section{Monte Carlo Calculations}

\subsection{Trial Wave Function}

In all QMC methods mentioned above, a good trial function is very
important for accurate results.
The convergence time in a released-node calculation
can be reduced with a better trial function
while the nodes of a trial function determine
the ultimate accuracy of a fixed-node calculation.
The usual choice of a trial function is of the Slater-Jastrow type
 \begin{equation}
    \Psi_{T}(R) \: = \: \det(\varphi_{mn}) \; \exp
[-\sum_{i<j}^{N}u(r_{ij})] \; ,
 \end{equation}
where $\varphi_{mn}=e^{i{\bf k}_{m}\cdot{\bf r}_{n}}$
for a homogeneous liquid phase.
The nodes are determined by only the Slater determinant.
We use the two-body correlation function $u(r)$ that minimizes the variational
energy in the Random Phase Approximation \cite{rpa,cep78}.
With this trial wave function, 
Ceperley calculated
the ground-state properties of the electron gas, using
the VMC \cite{cep78}, the fixed-node \cite{conf} 
and the released-node DMC method \cite{alder}.

 In order to improve the nodes,
we consider a more  complicated trial function which includes
$backflow$  and $three$-$body$ correlations \cite{kwon93}.
Our wave function has the form of
 \begin{equation}
    \Psi_{T}(R)=\det (e^{i{\bf k}_{i}\cdot{\bf x}_{j}})\: \exp [-\sum_{i<
j}^{N}\tilde{u}(r_{ij})-\frac{\lambda_{T}}{2} \ 
\sum_{l=1}^{N}{\bf G}(l) \cdot {\bf G}(l)] \; ,
 \end{equation}
where ${\bf x}_{i}$'s are $quasiparticle$ coordinates defined as:
 \begin{equation}
 {\bf x}_{i}={\bf r}_{i}+\sum_{j \neq i}^{N} \eta(r_{ij}) \: ({\bf r}_{i}-
{\bf r}_{j}) \; ,
 \end{equation}
 \begin{equation}
  {\bf G}(l)=\sum_{i \neq l}^{N} \xi(r_{li}) \: ({\bf r}_{l}-{\bf r}_{i}) \; ,
 \end{equation}
and
 \begin{equation}
    \tilde{u}(r)=u(r)-\lambda_{T} \, \xi^{2}(r) \, r^{2} \; .
 \end{equation}
In addition to the two-body correlation, this trial function
includes the three-body correlation, ${\bf G}(l) \cdot {\bf G}(l)$,
and the state-dependent correlation, ${\bf k} \cdot
({\bf r}_{i}-{\bf r}_{j}) \eta(r_{ij})$,
which incorporates the hydrodynamic {\em backflow} \cite{bflow}.
We call $\xi(r)$
the ``three-body correlation function" and $\eta(r)$ the
``backflow correlation function".
Note that it is the backflow correlation which makes
the nodes of the wave function different from 
those of the Slater-Jastrow trial function.

 Our calculations are done for $N$ electrons in a cube
with periodic boundary conditions.
The Ewald method \cite{ewald} is used for
the Coulomb potential and 
the two-body correlation $u(r)$ to minimize size effects.
The higher-order correlation functions, $\eta(r)$ and $\xi(r)$, 
are required to
go to zero smoothly at a cutoff distance $r_{c}$ set to half
the side of the simulation cell we use:
 \begin{equation}
    f(r) \; \longrightarrow \; f(r)+f(2r_{c}-r)-2f(r_{c}) \; .
 \end{equation}
The backflow and the three-body correlation function are parametrized as
\begin{equation}
   \eta(r)=\lambda_{B} \frac{1+s_{B}r}{r_{B}+w_{B}r+r^4} \; ,
\end{equation}
and
\vspace{-.08in}
\begin{equation}
   \xi(r)={\rm exp}[-(r-r_{T})^{2}/w_{T}^{2}] \; .
\end{equation}
This functional form for $\eta(r)$ satisfies the long-range
behavior ( $\sim 1/r^3$ ) in three dimensions predicted
by the local-energy method of Ref. \cite{kwon93}. 
It should be noted that the optimized $\eta(r)$ goes to zero rapidly
at the edge of the simulation box.
Our three-body correlation has
the same form as used for liquid $^{3}$He in Ref. \cite{he2}.

 In order to optimize our higher-order correlation functions,
we minimize the variance of the local energy \cite{vari}, defined by
\begin{equation}
V_{\Psi_{T}} = \frac{\int dR \; \Psi_{T}^{2}(R) \; (\; E_{L}(R)-E_{v}\; )^{2}}
{\int dR \; \Psi_{T}^{2}(R)} \; .
\end{equation}
If our trial function $\Psi_{T}$ were an exact eigenfunction
of the Hamiltonian, the variance would be zero.
Because the variance is a non-linear function of the parameters
we cannot be certain that we have achieved converged results for this
class of trial functions.  

The optimum variational
parameters that we have obtained as a function
of density, are given in Table \ref{table1}.
Fig. (1) shows the effect (in the logarithm of the wave function) of two 
electrons
a distance $r$ apart coming from the three-body term. There is a very strong
density dependence. The effect is almost negligible at $r_s=1$ but as large
as 10\% at $r_s \geq 10$. Negative values imply that electron configurations
in which the ``forces'' (coming from $\xi(r)$) are not ``balanced'' are
slightly enhanced.
Fig. (2) shows the magnitude of the displacement 
of the quasiparticle coordinate
caused by an electron a distance $r$ away. 
This is an estimate of the distance that the
free-fermion nodal surfaces are displaced by backflow. 
The strongest effects are observed
when two electrons are very close, for distances less than the average nearest
neighbor distance which is $2$ in the units we have used.
We also note that the backflow potential is
attractive for $r_s \geq 10$. 
We expect that the displacement of the quasiparticle
coordinates is on the order of 0.01 $a$.
Assuming that this is the case,
the released-node calculation with 
a relatively short projection time should be
able to correct the nodal surfaces from those 
of a free fermion trial function.

The main difficulties in the use of the backflow wave function is that
firstly,
the implementation is considerably more complex than for the 
Slater-Jastrow form, and secondly, because update formulas cannot be used
to speed up single particle moves, Monte Carlo moves are of 
updating all particles simultaneously.
Details and fuller discussion of the algorithm are given in Ref. \cite{kwon93}.

\subsection{Ground State Energy}

We first calculated the ground-state energy of the system with $N=54$ electrons
at a density range of $1 \le r_s \le 20$ 
by both VMC and fixed-node DMC methods. 
Table \ref{table2} shows the results
obtained from the improved trial wave functions in Eq. (5)
as well as the Slater-Jastrow wave functions. It can be seen that 
both VMC and fixed-node 
calculations with the trial functions including backflow
correlation improve significantly the Slater-Jastrow results
at all densities considered. However, the three-body correlation
is found to have minimal effect for $r_s \le 5$, which corresponds
to typical metallic densities. 

Fig. (3) shows the effects of backflow and three-body correlations
on the correlation energy that is missed by the  Slater-Jastrow
wave function  both from the variational and the fixed-node calculation.
In the following discussion, our best results (the backflow fixed-node
energies) are assumed to be exact.
We will examine this assumption at the end of 
this section.
At high densities of $r_s \le 5$, the effect due to the three-body
correlation is negligible and the backflow effect is dominant.
However, as the density decreases, the three-body effect increases
while the backflow effect decreases. 
We can conclude from the trends of Fig. (3) that at the
density where Wigner crystallization occurs, estimated
to be $r_{s} \sim 100$ by Ceperley and Alder \cite{alder},
the effect in the energy of the three-body term will be much larger than the 
backflow term.
This is consistent with the expectation that
backflow correlation is energetically less important as electrons are localized
by strong correlation at low densities.
Note, however, that the actual effect of the backflow
correlation on the wave function 
decreases
with density, as shown in Fig. (2).

The combined effects of both higher-order correlations in the variational
wave function account for 60\% to $80 \%$ of the correlation
energy missing in the Slater-Jastrow function.
At high densities ($r_{s} \leq 5$),
this variational energy is shown to be roughly as good as
the Slater-Jastrow fixed-node DMC energy, 
which captures 70 to $80 \%$ of the missing correlation
energy throughout our density range.

The backflow and three-body effects in the electron gas
discussed above are very similar to
the situation in two dimensions. See Fig. (4) of Ref. \cite{kwon93}.
The only notable difference is that at the lowest density
considered ($r_s=20$), the backflow effect is 
more important than the static three-body correlation in three dimensions
while two correlations have virtually equal importance
in two dimensions.
This can be understood in terms of the increased importance
of correlations in lower dimensions for the same value of $r_s$; 
for example, this is 
reflected in the fact that Wigner crystalization
occurs at smaller $r_s$ in two dimensions than in three dimensions
\cite{tanatar}.

Fig. (4) shows the correlation energies missing from the Slater-Jastrow
wave function and from the three-body and backflow wave function divided
by the kinetic energy.
Since the kinetic energy operator does not commute
with the Hamiltonian, the kinetic energy cannot be computed
directly with the distribution $\phi_0(R) \Psi_T(R)$
sampled through the diffusion process in Eq. (3).
It has been estimated by making an extrapolation
between the VMC and the DMC results\cite{sign}.
It is clear from the figure that the missing correlation energy
from both types of trial wave functions becomes a smaller fraction
of the kinetic energy at higher densities.

Since our calculation has been done on the system with 
a finite number of electrons,
we extrapolate the energies 
to the thermodynamic limit to compare with other calculations.
We follow the extrapolation scheme based upon 
the Fermi liquid theory \cite{hydrogen,tanatar}, which assumes that the energy
per particle for a finite system with the periodic boundary condition
is related to the bulk energy by
\begin{equation}
  E_{N}=E_{\infty}+b_{1}(r_{s}) \Delta T_{N} + b_{2}(r_{s}) \frac{1}{N} \; .
\end{equation}
Here, $E_N$ ($E_{\infty}$) is the total energy per electron
of the finite (infinite) system and $\Delta T_N$ is the
free particle kinetic energy differences between two systems.
We determine the parameters $E_{\infty}$, $b_{1}$, and $b_{2}$ by
a least-squares fit to VMC calculations with Slater-Jastrow trial functions
at different values of $N=54,66,114,162,246$.
In Table \ref{table4} are shown the energies, fitted parameters, and
the $\chi^{2}$ value of the fit.
The reasonable values of $\chi^2$ show that the Fermi liquid theory
completely explains the size dependence of the energy to statistical
accuracy of the VMC energies over this range of particle numbers.
To extract the extrapolated three-body and backflow DMC energy for
the infinite system, $E_{\infty}^{3BF-DMC}$,
we did the DMC runs only at $N=54$ whose results
are shown in Table \ref{table2} and then
use the parameters determined from VMC to get
$E_{\infty}^{3BF-DMC}$.
It is assumed that the size dependences for the VMC (SJ)
and the DMC (3BF) results are the same. 
This assumption needs to be tested in future calculations.
The same procedure was successfully applied to assess the finite-size
effects in our previous QMC calculation for the two-dimensional electrons
\cite{kwon93}. 

One can see in Table \ref{table4} that our extrapolated backflow
fixed-node energies are lower, even if the differences are small, than
Ceperley and Alder's released-node results as well as Ortiz and
Ballone's Slater-Jastrow fixed-node results.
Our present results show that the calculations of Ceperley and Alder  
only got approximately half of the Slater-Jastrow fixed-node error with their 
released-node procedure due to computer limitations at that time.
Considering that a fixed-node energy is an upper bound
to the true ground-state energy, this validates our assertion that
our backflow fixed-node results are accurate.

Since the fixed-node results depend only on the
nodal structures of the trial functions used, one can speculate that
the nodal locations of the backflow wave function are 
fairly close to those of the exact ground state.
Without more investigation, we cannot quantify this statement, because there
is not a simple relationship between nodal locations 
and fixed-node energy.
The accuracy of the backflow nodes
was also shown in our previous  released-node (transient-estimate) calculation
for the two-dimensional electron gas \cite{kwon96}.

Although comparison with well-converged exact results is the best
method of assessing the accuracy of a fixed-node result for the energy,
in the remainder of this section we develop two other methods that require
only the VMC and fixed-node DMC energies.
Both methods rely on the fact that the 
errors in the variational energy $E_{VMC}$, the variance of the 
local energy $V$, and the fixed-node energy $E_{FN}$ should
all be quadratic in the difference between a trial function and the
true ground state. Thus, as a trial function is significantly improved in
going from a two-body level (Slater-Jastrow) to a three-body
level (backflow and three-body), 
one can estimate the exact energy by the relative improvements of 
the variational energy relative to the variance and the fixed-node energy.

The variances of the local energy (Eq. (12)) for the various
trial wave functions, are given in Table \ref{table2}
and plotted in Fig. (5) at $r_s=10$.
As can be seen,
the variance decreases roughly proportional to the drop in energy for the
four trial functions considered.
The dotted line in Fig. (5) represents a linear fit and the triangle our
best (backflow) fixed-node energy.
There is no fundamental reason why
the energy and variance for general trial functions would have a linear 
relationship.
However, in practice this relation is often observed \cite{kwon93}. 
The observed linear relationship
both validates our optimization procedure and provides an independent
estimate of the exact energy. 

Shown in Table \ref{table2} is our estimate of the error of the computed
backflow fixed-node energy obtained from the energies and variances of the 
best (backflow + three-body) and worst (Slater-Jastrow) 
trial functions, which is based on the following assumption:
\begin{equation}
\frac{V^{(k)}}{E_{VMC}^{(k)}-E_0} = constant.
\end{equation}
$\epsilon_V$ in Table II is the difference between
this extrapolation $E_0$ and our best fixed-node energy.
We extrapolated using only the results from the best 
and worst trial functions to minimize 
the extrapolation error.  
There is the Temple
lower bound \cite{temple} to the ground-state energy 
which involves the energy and the variance.
However, it is not useful for many-body systems. Because our procedure is not
rigorous, there is no guarantee that the estimate will lie below our computed
best fixed-node result. In fact at $r_s=10$ the estimate lies above it. 
Our next extrapolation
procedure does not have this problem.

In going from the two-body to three-body level, one can also assume that the
nodal positions improve at the same rate as the variational energy so that
we can assume:
\begin{equation}
\frac{E_{FN}^{(k)}-E_0}{E_{VMC}^{(k)}-E_0} = constant.
\end{equation}
Using this equation with our best and worst energies in both VMC and
DMC calculations,
we determine $E_0$ and hence the error 
in the backflow fixed-node energy is
shown as $\epsilon_{FN}$ in Table \ref{table2}.
Again this procedure has no fundamental validity since it is possible
to improve the variational energy without affecting the nodes 
by improving the bosonic correlations.
This estimate shows that considerably larger corrections might
be expected from exact calculations, from $0.6mRy$ at $r_s=1$ to 
$0.1 mRy$ at $r_s=20$.

It can be seen in Table \ref{table2} that 
the estimated fixed-node errors ($\epsilon_{V}$ and $\epsilon_{FN}$) 
are smaller at all densities considered
than the energy improvements due to the nodal change from
the Slater-Jastrow function to the backflow wave function.

\section{Conclusion}
We have studied the correlation energy of the interacting
three-dimensional electron gas, using 
VMC and fixed-node DMC calculations including
the three-body and the backflow correlation.
The additional correlation energy due to backflow is
dominant over the three-body effect in the high density regime
but the relative importance of the former decreases as the density
is reduced. This is the same trend as was found  for the two-dimensional
electron gas \cite{kwon93} except that the importance of backflow 
is more significant in higher dimensions, especially
at low densities.  This is due to the fact that in two dimensions the
effects of interactions are larger than in three dimensions at a given $r_s$
and other effects tend to dominate more over the effects of 
backflow.

The variational wave function with backflow and three-body correlations
is a large improvement over the Slater-Jastrow function.
We find that these higher-order correlations account for 60 to $80 \%$ 
of the remaining correlation energy beyond the Slater-Jastrow
variational results.
Since backflow changes the nodes, 
the fixed-node DMC results are also significantly improved.
The fixed-node method based upon the Slater-Jastrow nodes is found to capture
no more than $80 \%$ of the remaining correlation energy.

After making a careful finite-size analysis, 
we have compared our backflow fixed-node energies with 
Ceperley and Alder's released-node
results. These two independent calculations using different
methods are found to give nearly identical results within statistical
and systematic errors. From a linear extrapolation to zero variance 
of the local energy,
we find further evidence that our backflow fixed-node results
are very  close to the true ground-state energy.

For future work, we conclude that one
should be able to use the much improved wave functions,
better released-node methods \cite{caf}, with more size-dependence studies
and full utilization of current computer hardware to achieve an order
of magnitude more accurate results for the energy of the electron gas
than was done nearly two decades ago.

\section{Acknowledgement}

This work has been supported by 
the Korea Science and Engineering Foundation under grant 96-0207-045-2
and through its SRC program, and by the National Science Foundation under grant
DMR 94-224-96.

\pagebreak

\begin{figure}
\caption{ The three-body contribution to the logarithm of the wave function 
for a pair
of electrons separated by a distance $r$. Solid line, $r_s=1$; dotted, $r_s=5$;
dot-dashed, $r_s=10$; long dashed, $r_s=20$.}
\label{fig1}
\end{figure}

\begin{figure}
\caption{ The change in quasi-electron coordinate due to a pair
of electrons separated by a distance $r$. Solid line, $r_s=1$; dotted, $r_s=5$;
dot-dashed, $r_s=10$; long dashed, $r_s=20$.}
\label{fig2}
\end{figure}

\begin{figure}
\caption{Effects of {\em three-body} and {\em backflow} correlations
 as a function of the density of the system. The vertical axis shows
 $\Delta E/\Delta E_{SJ}=
  (E-E_{DMC}^{3BF})/(E_{V}^{SJ}-E_{DMC}^{3BF})$,that is,
 top axis corresponds to the Slater-Jastrow variational energy $E_{V}^{SJ}$ and
 bottom axis to the fixed-node DMC energy $E_{DMC}^{3BF}$,
 calculated with our best trial function including
 {\em three-body} and {\em backflow} correlations.
 The diamonds show the effect of only {\em three-body}
 correlation, the circles the effect of only {\em backflow}
 and the squares represent the combined effect of both correlations.
 Finally, the filled triangles show the result 
 using the fixed-node DMC method with
 free-fermion nodes of the Slater-Jastrow function.}
\label{fig3}
\end{figure}

\begin{figure}
\caption{The energy missing from the Slater-Jastrow wave function ($\diamond$)
and from the three-body and backflow wave function ($\bullet$) divided
by the kinetic energy as a function of the density parameter $r_s$.
The vertical axis shows $(E_V - E_{DMC}^{3BF})/<T>$.}
\label{fig4}
\end{figure}

\begin{figure}
\caption{Variational energy versus the variance of local energy at $r_s=10$.
Each point $\bullet$ represents one variational calculation: from higher
to lower energies, the Slater-Jastrow, three-body, backflow, and
(backflow $+$ three-body) results.
The filled triangle represents
our backflow fixed-node result 
and the dotted line shows
a linear fit through $\bullet$ points.
The statistical errors of the data are smaller than the sizes of the symbols.}
\label{fig5}
\end{figure}
\pagebreak

\begin{table}
 \caption{Optimized variational parameters of {\em three-body} 
    and {\em backflow } correlation functions for $N=54$.}

 \begin{tabular}{r|ccccccc}
  $r_{s}$  &  $\lambda_{B}$  &  $s_{B}$  &  $r_{B}$  &  $w_{B}$   &$\lambda_T$
    &  $r_{T}$  &  $w_{T}$  \\ \hline
    1.0     &  0.025          &  0.395    &  0.210    &  0.689    &   0.006
    &  0.293    &  0.949    \\
    5.0     &  0.105          &  0.158    &  0.180    &  0.670    &  -0.060
    &   0.286   &  1.176    \\
   10.0     &  0.959          &  -0.672   &  0.247    &  3.788    &  -0.258
    &   0.257   &  0.918    \\
   20.0     &  1.249          &  -0.938   &  0.275    &  3.787    &  -0.255
    &   0.252   &  0.911    \\
 \end{tabular}

\label{table1}
\end{table}

\vspace{.5in}

\begin{table}
  \caption{ VMC and fixed-node (FN) DMC 
energies $E$ and the variances of the local energy $V$ 
obtained with various trial
wave functions for $N=54$
(SJ: the Slater-Jastrow function,
3BD: three-body correlation, BF: backflow correlation).
The energies are in units of $Ry$ per electron and
the variances in units of $r_s^4 (Ry/{\rm electron})^2$.
Also shown are our estimations of the fixed-node error $\epsilon$
in the backflow fixed-node DMC calculation.}

  \begin{tabular}{ccccc}
           & $r_{s}=1.0$ & $r_{s}=5.0$ & $r_{s}=10.0$ & $r_{s}=20.0$ \\ \hline
 $E_{VMC}^{SJ}$  & 1.0669(6)  & -0.15558(7)  & -0.10745(2)  & -0.06333(1) \\ 
 $E_{VMC}^{SJ+3BD}$  & 1.0663(5)  & -0.15569(5)  & -0.10773(2)  & -0.06348(1) \\
 $E_{VMC}^{SJ+BF}$  & 1.0617(4)  & -0.15729(5)  & -0.10829(2)  & -0.06365(1) \\
 $E_{VMC}^{SJ+3BD+BF}$ & 1.0613(4) & -0.15735(5) & -0.10835(2) & -0.06378(2) \\ 
                   &            &            &           &          \\
 $E_{FN}^{SJ}$    & 1.0619(4)  & -0.15734(3)  & -0.10849(2)  & -0.06388(1) \\
 $E_{FN}^{SJ+3BD+BF}$ & 1.0601(2) & -0.15798(4) & -0.10882(2) & -0.06403(1) \\
                   &            &            &           &             \\
  $V^{SJ}$       & 0.0213(4)  & 0.0266(3)  & 0.074(1)  & 0.189(3) \\
  $V^{SJ+3BD}$    & 0.0205(4)  & 0.0229(4)  & 0.054(2)  & 0.144(3) \\
  $V^{SJ+BF }$    & 0.0054(3)  & 0.0069(2)  & 0.027(1)  & 0.111(2) \\
  $V^{SJ+3BD+BF}$ & 0.0053(2)  & 0.0066(2)  & 0.026(1)  & 0.079(2) \\
                   &            &            &            &             \\
  $\epsilon_V$     & 0.0007(6)  &-0.00005(8) & 0.00002(5) & 0.00007(4) \\ 
  $\epsilon_{FN}$  & 0.0006(4)  & 0.00036(8) & 0.00027(5) & 0.00013(3) \\
  \end{tabular}

\label{table2}
\end{table}

\begin{table}
  \caption{Size dependence in the Slater-Jastrow VMC method of normal electron
liquid at $1 \leq r_{s} \leq 20$ and $\chi^{2}$-fit parameters.
Also shown are the extrapolated DMC energies at an infinite system
($E_{\infty}^{SJ-DMC}$ and $E_{\infty}^{3BF-DMC}$), 
 Ceperley and Alder's released-node result (CA$^*$),
 and Ortiz and Ballone's Slater-Jastrow fixed-node result (OB$^{**}$).}

  \vspace{.2in}

  \begin{tabular}{cccccc}
    \multicolumn{2}{c}{ } & $r_{s}=1.0$ & $r_{s}=5.0$ & $r_{s}=10.0$ & 
$r_{s}=20.0$ \\ \hline 
               & $N=54$ & 1.0669(6) & -0.15558(7) & -0.10745(2) & -0.06333(1) \\
               & $N=66$ & 1.1496(5) & -0.15166(4) & -0.10637(2) & -0.06303(1) \\
  $E_{V}^{SJ}$ & $N=114$& 1.2079(5) & -0.14867(3) & -0.10552(2) & -0.06278(1) \\
               & $N=162$& 1.1162(4) & -0.15238(3) & -0.10642(1) & -0.06270(1) \\
               & $N=246$& 1.1938(3) & -0.14886(3) & -0.10548(1) & -0.06275(1) \\
               &        &           &             &             &             \\
    \multicolumn{2}{c}{$E^{SJ-VMC}_{\infty}$}& 1.1795(4) & -0.14914(3) & 
-0.10549(2) & -0.06273(1) \\
    \multicolumn{2}{c}{$b_1(r_s)$} & 1.096(6) & 1.18(1) & 1.21(2) & 1.22(3) \\
    \multicolumn{2}{c}{$b_2(r_s)$} & -1.16(5)  & -0.134(4) & -0.051(2) & 
-0.0181(7)\\
    \multicolumn{2}{c}{$\chi^2$}   & 1.20 & 1.29 & 2.22 & 2.26 \\
               &        &           &             &             &             \\
    \multicolumn{2}{c}{$E^{SJ-DMC}_{\infty}$}  & 1.1744(4) & -0.15094(4) &  
-0.10654(2) & -0.06329(1) \\
    \multicolumn{2}{c}{$E^{3BF-DMC}_{\infty}$} & 1.1726(2) & -0.15158(5) & 
-0.10687(2) & -0.06344(1) \\
    \multicolumn{2}{c}{CA$^*$}  & 1.174(1) & -0.1512(1) & -0.10675(5) & 
-0.06329(3) \\
    \multicolumn{2}{c}{OB$^{**}$}  & 1.181(1) & -0.1514(3) & - & -  \\
  \end{tabular}

\label{table4}
\end{table}

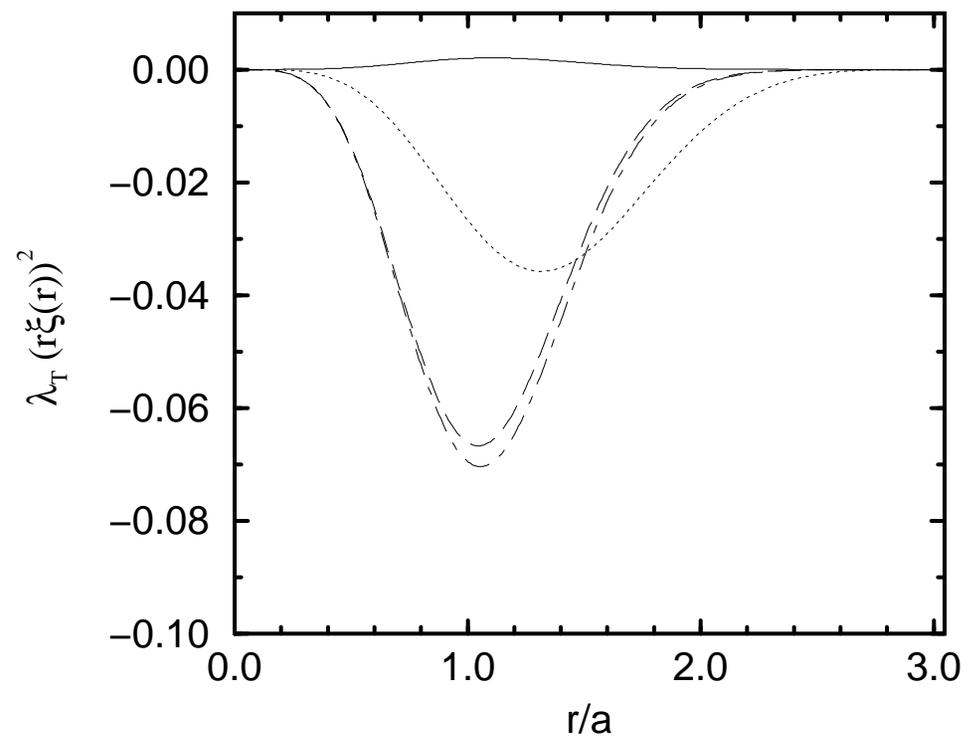
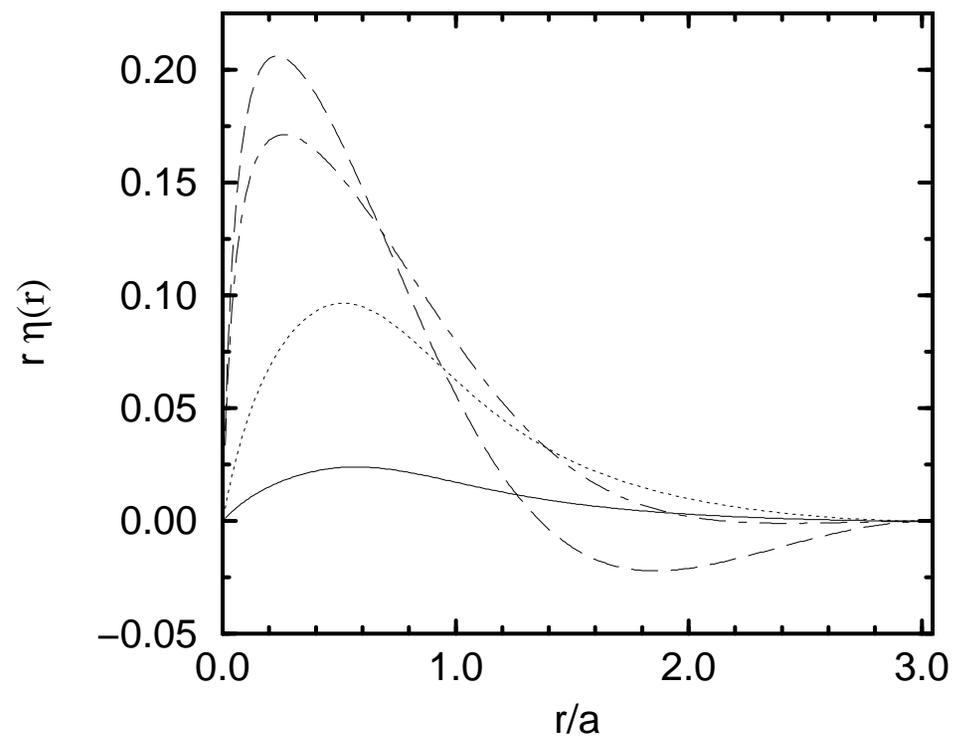
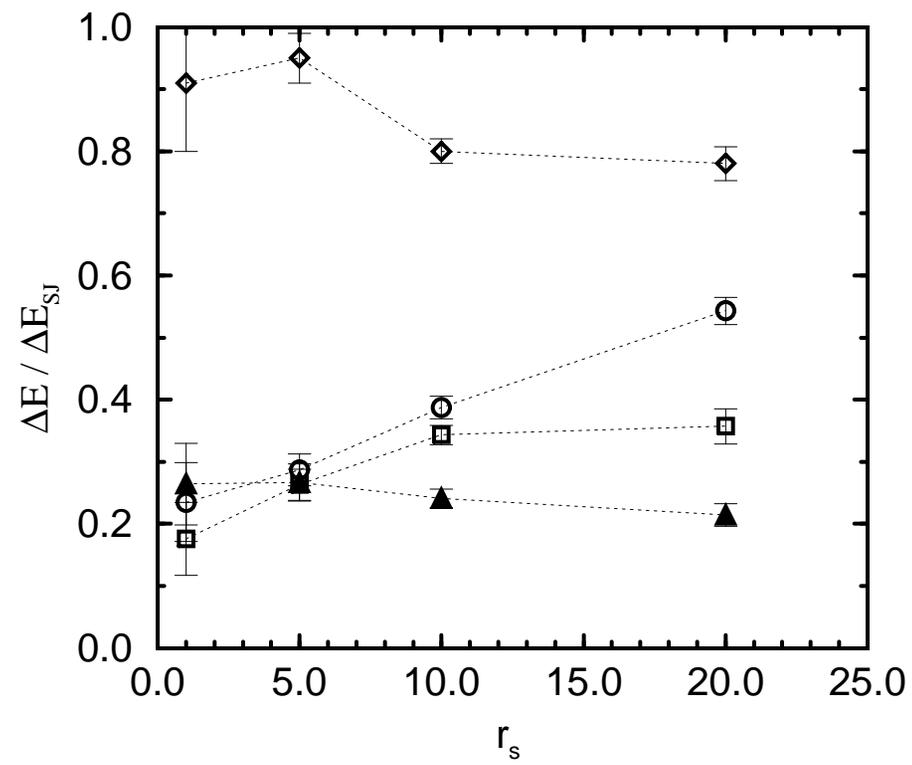
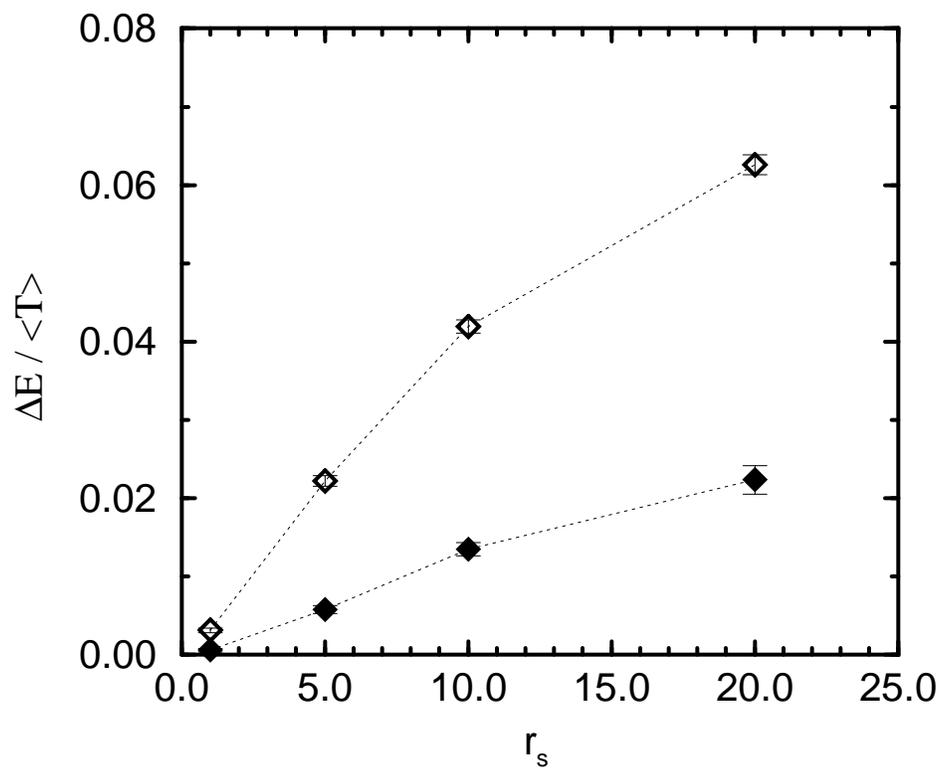
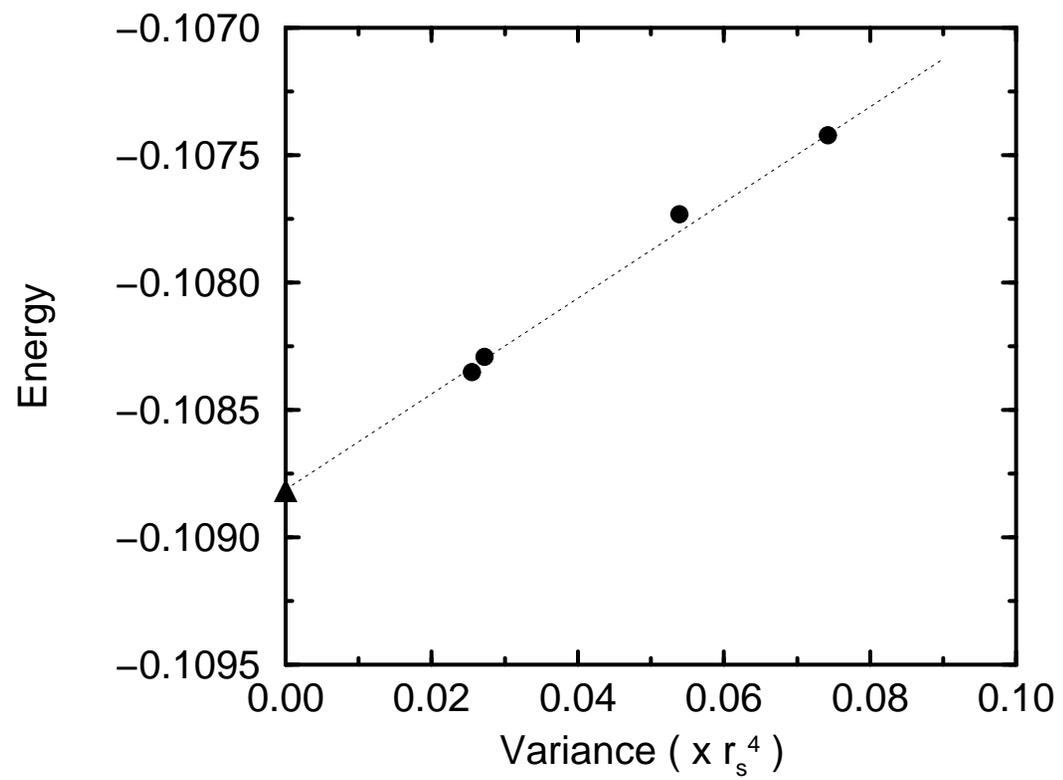
\end{document}